\documentclass[12pt]{amsart}
\usepackage{amsmath}
\usepackage{amsthm}
\usepackage{amssymb}
\usepackage{cite}

\newcommand{\tr}{\mathop\mathrm{tr}}
\newcommand{\rk}{\mathop\mathrm{rk}}
\newcommand{\dd}{d}
\newcommand{\ks}{k}

\newcommand{\up}[1]{^{(#1)}}

\def\E{{\mathcal E}}
\def\H{{\mathcal H}}
\def\K{{\mathcal K}}

\def\V{{\mathcal V}}

\newtheorem{thm}{Theorem}[section]

\newtheorem{cor}[thm]{Corollary}

\newtheorem{lemma}[thm]{Lemma}

\newtheorem{defn}[thm]{Definition}
\theoremstyle{remark}
\newtheorem{remark}[thm]{Remark}
\newtheorem{remarks}[thm]{Remarks}

\newtheorem{ex}[thm]{Example}

\newcommand{\CC}{\mathbb C}

\begin{document}

\title[Decoherence-Insensitive Quantum
Communication]{{Decoherence-Insensitive Quantum Communication by
Optimal {\boldmath $C^*$}-Encoding}}
\author[B.~G.~Bodmann, D.~W.~Kribs, V.~I.~Paulsen]{Bernhard G. Bodmann$^1$,  David W. Kribs$^{2,3}$, and Vern I. Paulsen$^4$}

\address{$^1$Department of Applied Mathematics, University of Waterloo,
Waterloo, Ontario N2L 3G1, Canada}
\address{$^2$Department of Mathematics and Statistics,
University of Guelph, Guelph, Ontario  N1G 2W1, Canada}
\address{$^3$Institute for Quantum Computing, University of Waterloo, Waterloo,
Ontario N2L 3G1, Canada}
\address{$^4$Department of Mathematics, University of Houston, Houston, TX 77204-3476, USA}


\begin{abstract}
The central issue in this article is to transmit a quantum
state in such a way that after some decoherence occurs, most of
the information can be restored by a suitable decoding operation.
For this purpose, we incorporate redundancy by mapping a given
initial quantum state to a messenger state on a larger-dimensional
Hilbert space via a $C^*$-algebra embedding. Our noise model for
the transmission is a phase damping channel which admits a
noiseless or decoherence-free subspace or subsystem. More
precisely, the transmission channel is obtained from convex
combinations of a set of lowest rank yes/no measurements that
leave a component of the messenger state unchanged.
The objective of our encoding is to
distribute quantum information optimally across the
noise-susceptible component of the transmission when the noiseless
component is not large enough to contain all the quantum
information to be transmitted. We derive simple geometric
conditions for optimal encoding and construct examples.
\end{abstract}

\date{\today}
\maketitle

\section{Introduction}\label{S:intro}

The use of quantum systems as a medium to carry information
enables algorithms and communication techniques not possible in
the purely classical setting. While there is considerable
excitement about these new capabilities, their implementation
poses severe theoretical and experimental challenges. 
One of the major impediments to
large-scale experimental advances
is the decoherence of
quantum systems as they evolve in time. Indeed, finding efficient
methods for ``quantum error correction'' is of central importance
in the fields of quantum computation, cryptography and communication.

A fundamental, ``passive'' technique for quantum error correction
grew out of the recognition that certain error models contain
symmetries that can be used to hide encoded qubits from the
overall noise of the model
\cite{PSE96,DG97c,ZR97c,LCW98a,KLV00a,Zan01b,KBLW01a}. This scheme
relies on the identification of what are now known as noiseless or
decoherence-free subspaces or subsystems. (For convenience we will
just use the noiseless terminology.) A number of experimental
advances have been based on this approach
\cite{KBAW00,KMRSIMW01,FVHTC02,VFPKLC03}. Noiseless subsystems and
subspaces have recently received wider attention, such as in
quantum information processing and cryptography
\cite{BRS03,BGLPS04,WZL05}, and in quantum cosmology and gravity
\cite{DMS04,KM05,BMS06}. Moreover, a structure theory is now
emerging \cite{KLP05,KLPL06,LS05,NP05,CK06,Kni06a} which
shows, among other things, how such subsystems and subspaces can
be identified \cite{CK06,Kni06a}.

The basic problem we address in this paper is, given a certain
class of decoherence error models, how to distribute quantum
information optimally across the noise-susceptible component of
the transmission when the noiseless component is not large enough
to encode all the information to be transmitted. 
Our analysis is motivated by recent
work \cite{SH03,HP04,BP05,XZG05,Kal06} on the construction and use
of frames to incorporate redundancy in classical signal
transmission.

For the purposes of this paper, a quantum communication
framework is given by triples from a collection of quantum channels.
Each triple contains the maps for encoding,
transmission and reconstruction of quantum states.

We exclusively consider encoding maps that are  trace-preserving
$C^*$-alge\-bra embeddings in order to preserve the structure of
the set of encoded quantum states. We call such maps
$C^*$-encodings. Consequently, the encoding and the subsequent
reconstruction are quantum channels, that is, completely positive
and trace-preserving (or at least non-increasing for the trace).
The encoding channel incorporates redundancy by mapping a quantum
state on a given Hilbert space to a ``messenger'' state on a
larger-dimensional Hilbert space.

It is common in the context of quantum error correction to model
noise in transmissions by a so-called interaction algebra
$\mathcal A$ and search for states that are in the commutant of
$\mathcal A$. The interaction algebra we consider is given by the
commutative set of operators $\mathcal A = \{D \otimes I: D \mbox{
diagonal}\}$ on a Hilbert space ${\mathbb C}^m \otimes {\mathbb
C}^l$, admitting a noiseless subsystem in the second component.
However, we do not consider the entire commutative interaction
algebra $\mathcal A$, but rather focus on a ``minimal'' set
$\mathcal{Q}\up 1$ of error models for the transmission. This set
contains all nontrivial lowest-rank projective measurements and
their convex combinations. Such errors model generalized phase
damping, that is, decoherence due to minimal interaction with a
macroscopic system that performs with certain probabilities one
among several possible yes/no measurements on our system. The
commutativity means that the outcomes of the measurements obey the
rules of classical Boolean logic. Apart from naturally occurring
decoherence due to interactions with an environment, the
motivation for considering this type of error is that such
measurements may be used to remove corrupted information in a
quantum state after the occurrence of a quantum error has been
detected.

Minimal projections in the interaction algebra
$\mathcal A$ are tensor products of diagonal matrix units $\{E_{jj}\}_{j=1}^m$
and the identity, henceforth denoted as $Q_j = E_{jj} \otimes I$.
Accordingly, a transmission channel which performs a minimal projective
measurement of a messenger state $M$
is $\mathcal{E}_j: M \mapsto Q_j M Q_j + Q_j^\perp M Q_j^\perp$,
where $Q_j^\perp$ complements the orthogonal projection $Q_j$ to the identity,
$Q_j + Q_j^\perp = I$. We denote the phase damping channels
obtained by convex combinations of such measurements as
\begin{equation}\label{transmissionmodel}
\mathcal{Q}\up 1= \Bigl\{ \sum_{j=1}^m p_j 
\mathcal{E}_j(M) \;\big| \mbox{ all } p_j \ge 0, \sum_{j=1}^m p_j =1  \Bigr\} \, .
\end{equation}

In the following, we shall use $B({\mathbb C}^d)$ to
denote the
algebra to be encoded. This algebra is embedded via the
encoding map $\Phi$ into the higher-dimensional algebra $B({\mathbb
C}^m\otimes{\mathbb C}^l)= B({\mathbb C}^m)\otimes B({\mathbb
C}^l)$. The subsequent decoding is a ``blind'' reversal of the
encoding operation.

For our purposes, an optimal $C^*$-encoding $\Phi$ minimizes
$\sup_\mathcal{E} e(\Phi,\mathcal{E})$, the supremum  of the Hilbert-Schmidt norms
for the reconstruction error
occurring among all quantum states and all noisy transmission channels $\mathcal E$
in the convex set $\mathcal{Q}\up 1$. While other measures for distortion are also
used in quantum information, we have chosen the Hilbert-Schmidt norm because it is
most closely related to the geometry of trace-preserving
$C^*$-algebra embeddings.

Our main result is that if $l<d$, a $C^*$-encoding $\Phi$ satisfies
\[
   \sup_{\mathcal{E} \in \mathcal{Q}\up 1} e(\Phi,\mathcal{E})/d
   \ge \left\{ \begin{array}{ll} 2-2d/ml \, ,& d/ml\in [0, 1/2] \, ;\\
     (8 m l d - 2 m^2 l^2 - 4 d^2)^{-1/2} \, ,& d/ml \in (1/2,1] \, .
\end{array}\right. 
\]
In addition, we characterize the case of equality as that for
which the encoding map is constructed from what we call a family of
``uniformly weighted projections resolving the
identity''. In many cases we show how such families can be
constructed as an extension of recent work on classical signal
encoding problems \cite{HP04}. However, we leave the problem of
determining the possible existence of these families in complete
generality for investigation elsewhere. Finally, for all 
$C^*$-encodings that are optimal for $\mathcal{Q}\up 1$, we derive 
an inequality for transmission channels with next-to-minimal decoherence 
and characterize the case of equality
as that for which all pair sums of the underlying projections 
have equal operator norm. This geometric condition on the ranges
of the projections
generalizes the notion of equiangular lines.

This paper is organized as follows: In
Section~\ref{S:communication} we describe the
concept of a quantum communication framework and examine 
the structure of this framework in the presence of a noiseless 
subsystem. In Section~\ref{S:UWP}, we obtain precise
estimates on the reconstruction error due to a noisy transmission and use 
this to characterize optimal encoding in terms of 
uniformly weighted projective resolutions of the identity. 
Section~\ref{S:ONM} treats optimality for secondary decoherence errors in the 
uniform encoding case.  Finally, we comment on a generalization
that includes the noiseless subspace case, with similar results 
on the reconstruction errors and optimal encoding schemes.


\section{Quantum Communication Framework}\label{S:communication}

We recall that for the purposes of this paper, a quantum communication framework
is given by three quantum channels: encoding, transmission, and decoding.
Hereby, a \textit{quantum channel} is understood to be a completely
positive, trace preserving (or at least trace non-increasing)
linear map between $C^*$-algebras.
We point the reader to
\cite{NieChu00,Pau02,BenZyc06} as entrance points into the
literature on completely positive maps and their use in quantum
computation and information.

\subsection{{\boldmath$C^*$}-Encoding and Decoding}\label{sS:ed}

Apart from being quantum channels, we require that these maps
preserve the algebraic structure of the quantum states to be
encoded.

\begin{defn}\label{def:encode/decode}
Let $\mathcal H$ and $\mathcal K$ be finite-dimensional Hilbert spaces.
We say that a map $\Phi: B(\mathcal{H}) \rightarrow B(\mathcal K)$
is a $C^*$-encoding of $B(\mathcal H)$ if $\Phi$ is a  trace-preserving
$*$-monomorphism. The associated decoding map is the Hilbert-Schmidt dual
$\Phi^* : B(\mathcal{K})\rightarrow B(\mathcal{H})$.
\end{defn}
\begin{remarks}
Algebra embeddings of the type introduced here 
have also been used for a unified treatment of 
a hybrid quantum/classical memory \cite{Kup03}.

In the context of our quantum communication framework,
the most important property of the encoding is that
it preserves the Hilbert-Schmidt inner product.
Thus $\Phi^*$ is a left inverse of $\Phi$,
and
encoding followed by decoding is the identity map, $\Phi^*
\Phi=\mathrm{id}$. Moreover, the dual $\Phi^*$  is completely
positive and non-increasing for the trace.

Another consequence of $\Phi$ being a Hilbert-Schmidt isometry is
that, instead of performing a measurement after decoding a
transmitted state, one may simply encode the measurement and apply
it directly to the transmitted state. 

Furthermore, the benefit of encoding with a $C^*$-algebra embedding 
is that a quantum algorithm performed on the initial state,
implemented by conjugation with a unitary operator, 
could be performed either prior
to encoding or afterwards.
\end{remarks}

\subsection{Error Model for Transmission}\label{sS:trans}

The transmission of an encoded state $M \in B({\mathcal K})$
is described by a quantum channel. When $\dim \mathcal{K} <
\infty$, such a map has a representation of the form
$\E(M) = \sum_a E_a M E_a^*$
for a finite set of operators $\{E_a\}\subset B(\H)$.
These operators $\{E_a\}$ are
said to form an ``error model'' for the channel. It is the impact
of such error operators that must be overcome to safely transmit
quantum information through the channel $\E$.

As discussed in the introduction (see
Eq.~(\ref{transmissionmodel})), here we consider phase damping
channels given by convex combinations of commuting 
projective measurements.

\begin{defn}
We call a quantum channel $\mathcal E$ a \textit{generalized phase-damping channel}
on $B({\mathcal K})$
if there is a commutative set of projections 
$\{Q_j\}_{j=1}^m \subset B(\mathcal{K})$ and a
probability vector $p \in \mathbb{R}_+^m$, $\sum_j p_j = 1$, such that
for $M \in B(\mathcal{K})$,
$$
   \mathcal{E}(M) = \sum_{j=1}^m p_j (Q_j M Q_j + Q_j^\perp M Q_j^\perp)\, ,
$$
where each $Q_j^\perp$ is the projection on the orthogonal complement
of the range of $Q_j$.
\end{defn}

We define the overall reconstruction error of the framework as
follows.

\begin{defn}
Let $\Phi: B(\mathcal{H}) \to B(\mathcal{K})$ be a $C^*$-encoding as in
Definition~\ref{def:encode/decode}, and let $\mathcal{E}:
B(\mathcal{K}) \to B(\mathcal{K})$ be a
completely positive, trace preserving map on the Hilbert space
$\mathcal K$. We then define the \textit{reconstruction error}
caused by $\E$ in the course of transmitting a state $W\in B(\mathcal{H})$ as
\begin{align*}
     Y    &= W - (\Phi^*\circ\E\circ\Phi)(W)
\end{align*}
and the maximum of the Hilbert-Schmidt norm of $Y$ over the set of
all states $\{W\ge 0$, $\tr W = 1\}$ as the \textit{worst-case error norm}
$$
    e(\Phi,\mathcal{E}) = \max_W (\tr[ Y Y^* ])^{1/2} \, .
$$
\end{defn}

\begin{remark}
The Hilbert-Schmidt norm is not directly motivated by informa\-tion-theoretic
considerations. However, in the situation considered here, it is a natural
way to quantify the impact of an error $\E$ in the transmission with $\Phi^*
\circ \E \circ \Phi$. In fact, the Hilbert-Schmidt norm is
non-increasing under any generalized phase-damping channel $\mathcal{E}$. 
This follows from the
convexity of the norm and from the case
of a single projection $Q$ and a state $M$,
\begin{multline} \nonumber
   \tr[(QMQ + Q^\perp M Q^\perp)^2 ] = \tr[ Q M Q^2 M Q + Q^\perp M (Q^\perp)^2 M Q^\perp]\\
   \leq \tr[ Q M^2 Q + Q^\perp M^2 Q^\perp] = \tr[ M^2 ] \, .
\end{multline}
Since this norm is also non-increasing when $\Phi^*$ is applied to
$\E(M)$, we can use it to measure the amount of corruption caused by the
error $\E$ and the subsequent decoding. When calculating the
Hilbert-Schmidt norm of the maximal reconstruction error, we thus
bound the maximal extent of corruption for any state.
\end{remark}

\subsection{Encoding in the Presence of Noiseless Subsystems}\label{S:ns}

Consider a Hilbert space $\mathcal K$ for a quantum system and a channel
$\E : B(\K)\rightarrow B(\K)$. Suppose we have a decomposition $\K
= (\K\up 1\otimes\K\up 2)\oplus \V$, so $\V = (\K\up 1\otimes\K\up
2)^\perp$, and for all states $W\up 1$ and $W\up 2$ on $\K\up 1$
and $\K\up 2$ there is a state $S\up 1$ on $\K\up 1$ with
\begin{equation*}
\E(W\up 1\otimes W\up 2\oplus 0) = S\up 1 \otimes W\up 2 \oplus 0\, .
\end{equation*}
Then the second component of the tensor product is said to be a
{\it noiseless subsystem} (or {\it noiseless subspace} in the case
$\dim\K\up 1=1$) for $\E$. The terminology ``decoherence-free''
subspace and subsystem is also quite common. This definition
ensures that quantum information encoded in the second component
of the tensor product is unaffected by the noise of the channel
$\E$.


For succinctness, we only treat the case $\V = \{0\}$ here. Remark~\ref{rem:gen}
sketches
how the results generalize to the case
$\mathop\mathrm{dim} \V > 0$.

For our transmission channels, the second component of the
tensor product decomposition ${\mathbb C}^N = {\mathbb C}^m
\otimes {\mathbb C}^l$ used for the transmission is noiseless,
whereas the first component may lose its off-diagonal entries. If
$l \ge d$, then we could simply encode $W$ as $M=\frac{1}{m}
I\otimes W$, which is unaffected by any decoherence in the first
component. So let us assume $l < \dd$, which means the dimension
of the second component is too small to contain all the
information given by an arbitrary state $W$ on ${\mathbb C}^d$ .

\subsubsection{$C^*$-encodings and Positive Operator Valued Measures}

Let us describe a useful concrete description of $C^*$-encodings
in terms of associated factorizations of POVMs.

\begin{lemma} \label{lem:coordop}
Given a $C^*$-encoding $\Phi: B({\mathbb C}^d) \to B({\mathbb C}^m\otimes {\mathbb C}^l)$,
then there exists a set of so-called coordinate operators
$\{V_j: {\mathbb C}^d\rightarrow{\mathbb C}^l\}$ resolving the identity $\sum_{j=1}^m V_j^* V_j = I$
such that $\Phi$ has the form
\begin{eqnarray*}
 \Phi: B({\mathbb C}^d) &\to& B({\mathbb C}^m\otimes {\mathbb C}^l) = B(\oplus_{j=1}^m {\mathbb C}^l)\\
          F &\mapsto& G = (G_{ij})_{i,j=1}^m, G_{ij}=  V_i W V_j^*.
\end{eqnarray*}
Moreover, the action of the decoding map is given by
\begin{equation}
G  \mapsto \Phi^*(G) =
\sum_{i,j=1}^m  V_i^* G_{ij} V_j \, .
\end{equation}

\end{lemma}
\begin{proof}
As $\Phi$ is a trace-preserving $C^*$-monomorphism of a simple,
finite-dimensional $C^*$-algebra $B({\mathbb C}^d)$, it follows
from the representation theory for such algebras \cite{Dav96} that
there is an isometry $\widetilde{V}:{\mathbb
C}^d\rightarrow{\mathbb C}^m\otimes{\mathbb C}^l = \oplus_{j=1}^m
{\mathbb C}^l$, $\widetilde{V}^* \widetilde{V} = I$, which
for all $F \in B({\mathbb C}^d)$ gives
\begin{equation}\label{phiequivform}
\Phi(F) = \widetilde{V} F \widetilde{V}^* \, .
\end{equation}
We may view $\widetilde{V}$ as a column matrix of coordinate
operators $V_j:{\mathbb C}^d\rightarrow{\mathbb C}^l$, $1\leq j
\leq m$,
\begin{equation}\label{Vtilde}
\widetilde{V} = ( \begin{matrix} V_1 \,\, V_2 \,\, \cdots \,\, V_m
\end{matrix})^t \,:\,{\mathbb C^d} \longrightarrow \oplus_{j=1}^m
{\mathbb C}^l \,; \quad f \longmapsto ( \begin{matrix} V_1f \,\,
V_2f \,\, \cdots \,\, V_m f \end{matrix})^t.
\end{equation}
The linearity of $\widetilde{V}$ ensures that each $V_j$ is
linear. Also note that ${\rm rank}(V_j)\leq l$, $1\leq j \leq m$.

Thus, the action of $\Phi$ is given by $G = \Phi(F) =
\widetilde{V} F \widetilde{V}^*$.  We may view $G$ as an operator
on ${\mathbb C}^m\otimes{\mathbb C}^l$ expressed in block matrix
form
\begin{equation}
G=(G_{ij})_{i,j=1}^m \mbox{ with } G_{ij}=V_i W
V_j^* \, .
\end{equation}
Now the Hilbert-Schmidt dual of $\Phi$ can be identified
as
\begin{equation*}
G  \mapsto \Phi^*(G) = \tilde V^* G \tilde V =
\sum_{i,j=1}^m  V_i^* G_{ij} V_j \, .
\end{equation*}
\end{proof}

\begin{remarks}
The operators $\{V_j\}$ define a ``positive operator valued
measure'' (POVM), $\{A_j\}$, where $A_j = V_j^*V_j$ and
\begin{equation}\label{povm}
I = \widetilde{V}^*\widetilde{V} = \sum_{j=1}^m A_j.
\end{equation}
Conversely, if we start with a POVM on ${\mathbb C}^d$ given by
$\{A_j\}$ such that $A_j\geq 0$, each $A_j$ has rank at most $l$,
and $\sum_{j=1}^m A_j=I$, upon factoring $A_j = V_j^*V_j$ one can
define $\widetilde{V}$ as in Eq.~(\ref{Vtilde}) and a
$C^*$-encoding as in Eq.~(\ref{phiequivform}).

Hence the $C^*$-encodings of Definition~\ref{def:encode/decode}
are completely characterized by factorizations of POVMs in this
way, and we shall use this concrete form in the analysis below. A
special case is of central importance in this paper.
\end{remarks}

\begin{defn}
Let $\mathcal H$ be a $d$-dimensional complex Hilbert space and
let $\{A_j\}_{j=1}^m\subset B({\mathcal H})$ be a POVM on $\H$. If
each operator is given by $A_j=k_j P_j$ with a non-negative
coefficient $k_j$ and a projection operator $P_j=P_j^* P_j$, then
we call $\{A_j\}$ a set of \textit{weighted projections resolving
the identity}. If $k_j = k = \dd/\sum_{j=1}^m \tr P_j$ for all $j
\in \{1, 2, \dots ,m\}$, we call $\{k P_j\}_{j=1}^m$  a set of
\textit{uniformly weighted projections resolving the identity}.
\end{defn}

\begin{remarks}
(i) Suppose all operators in a POVM $\{A_j\}_{j=1}^m$ have a
maximal rank $l$, and we denote $P_j$ the projection onto the range
of each $A_j$, and abbreviate the operator norms $k_j =\|A_j\|$.
Then from $k_j P_j \ge A_j$ and from taking traces on both sides
of Eq.~(\ref{povm}), we have
\begin{equation} \label{ineq:tr}
l \sum_{j=1}^m k_j \ge \sum_{j=1}^m \tr A_j = d \, .
\end{equation}
Equality holds in this inequality if and only if $A_j=k_j P_j$,
and $P_j$ has rank $l$ for
all $1\leq j \leq m$.

(ii) Weighted projections giving a resolution of the identity 
and their robustness against perturbations have
been investigated by Casazza and Kutyniok in the context of frames
and distributed processing, where they are known as
\textit{Parseval frames for subspaces} \cite{CK04} or
\textit{fusion frames} \cite{CKL06}. The $C^*$-encoding
derived from this special case of POVMs 
is of central importance in this paper.

(iii) If  $\sum_{j=1}^m \tr P_j > \dd$, then the ranges of the
$P_j$'s are not mutually orthogonal. However, this does not imply
that the ranges of the $P_j$'s have to intersect nontrivially, see
the example below.

(iv) Choosing orthonormal bases for the range of each $P_j$ and
scaling the basis vectors spanning the range of each $P_j$ by a
factor $\sqrt{k_j}$ would yield a Parseval frame of
$N=\sum_{j=1}^m \tr P_j$ vectors for the space ${\mathbb C}^\dd$.
If the weights are uniform, so is the frame.

(v) Given a set of weighted projections resolving the identity
such that $\sum_{j=1}^m \tr P_j > \dd$, we can replace each $P_j$
by $P_j'=I-P_j$ and each coefficient $k_j$ by
$k_j'=k_j/(\sum_{n=1}^m k_n - 1)$ in order to arrive at a
\textit{complementary resolution of the identity}, $\sum_{j=1}^m
k_j' P_j'=I$.
\end{remarks}

\subsubsection{Examples} \label{sec:examples}

We sketch the construction of
uniformly weighted projections resolving the identity.

\begin{ex}\label{ex1}
It is fairly straightforward to construct examples of weighted
projections resolving the identity for which all $\{k_j
P_j\}_{j=1}^m$ are rank-one and $m>d$. To this end, we take a
rank-$d$ orthogonal projection $G=G^* G$ on ${\mathbb C}^m$ and
factor it as $G=V V^*$, where $V$ is an isometry $V: {\mathbb C}^d
\to {\mathbb C}^m$. Now denote the $m$ column vectors of the
matrix representation for $V^*$ in the standard basis as
$\{f_j\}_{j=1}^m$, choose each $P_j$ to be the projection on the
one-dimensional subspace of ${\mathbb C}^d$ spanned by $f_j$, and
let the set of weights be $\{k_j = \|f_j\|^2\}$. It is then
straightforward to verify that $\sum_{j=1}^m k_j P_j = I$.

Holmes and Paulsen \cite[Remark 1.1]{HP04} show how to
make the weights uniform by rotating amongst the frame vectors.
\end{ex}

The next example describes how rank-one projections may be used
to construct higher-rank weighted projections which resolve the identity.

\begin{ex}\label{ex2}
If the Hilbert space is a tensor product ${\mathbb
C}^d={\mathbb C}^{l} \otimes {\mathbb C}^{q}$, so $d=l q$, and there
is a set of weighted rank-one projections
 $\{k_j \Pi_j\}_{j=1}^m$ resolving the identity on the second component ${\mathbb C}^q$, then
$\{k_j P_j = k_j I \otimes \Pi_j\}$ is a set of weighted rank-$l$
 projections resolving the identity on ${\mathbb C}^d$.

 If all weights are uniform in the resolution of the identity for the second component,
 the same holds for the tensor product.
 \end{ex}

However, there are examples of weighted projections realized on
tensor product spaces that are not of the type $\{I \otimes \Pi_j
\}$.

\begin{ex}\label{ex3}
We now describe a set $\{ k P_1,..., k P_m \}$ of $m$ uniformly
weighted projections resolving the identity in the case that
$d=2l$ is even.

This example has the property that $\rk(P_j)=l$
for every $j$ and if $i \neq j$, then the ranges of $P_i$ and
$P_j$ intersect only in the zero vector.

To establish the claimed properties, we identify in the usual way
$B(\CC^d)= B(\CC^l \oplus \CC^l)= B(\CC^2) \otimes B(\CC^l);$ that
is, as  $2 \times 2$ block matrices, each of whose entries is an
$l \times l$ matrix. Let
\begin{equation}
P= \begin{pmatrix} I & 0\\0 & 0 \end{pmatrix} \quad\quad {\rm and}
\quad\quad  U_j=
\begin{pmatrix} C_j & -S_j\\S_j & C_j \end{pmatrix} 
\end{equation}
for all $ 1 \le j \le m$,
where each $C_j$ and $S_j$ is an $l \times l$ diagonal matrix of
the form
\begin{equation}
C_j = \sum_{i=1}^l  \cos(\theta_{i,j})E_{ii} , \quad \quad S_j=
\sum_{i=1}^l \sin(\theta_{i,j})E_{ii},
\end{equation}
where $\theta_{i,j}= \frac{\pi (i-j)}{m}, 1 \le j \le m, 1 \le i
\le l,$ and $\{E_{ii}\}_{i=1}^l$ are the diagonal matrix units.
Using these matrices, we specify our set of $m$ projections of
rank $l$, by setting
\begin{equation}
P_j= U_j^*PU_j= \begin{pmatrix} C_j^2 & -C_jS_j\\ -C_jS_j
  & S_j^2 \end{pmatrix}, \quad 
\end{equation}
for all $ 1 \le j \le m$, so that $\rk(P_j)=l$ for all $j\in \{1,2, \dots m\}$.

Note that in this case, $k=d/\sum_j \tr(P_j)=d/(ml)= 2/l.$ To
verify that $\sum_{j=1}^m P_j= \frac{m}{2} I_d,$ observe that
$\sum_{j=1}^m \cos^2 ( \frac{\pi(j-i)}{m})= \frac{m}{2},$ while
$\sum_{j=1}^m \cos(\frac{\pi(j-i)}{m})$ $\sin(\frac{\pi(j-i)}{m})
= 0.$ Finally, to verify that the ranges of a pair $P_i$ and
$P_j$, $ i \neq j$, do not intersect nontrivially, we show $\|P_i
+ P_j \| < 2$. Since the operator norm is invariant under
conjugation with unitaries, we observe
\begin{equation}
\|P_i+P_j\| = \| P + U_iU_j^*PU_jU_i^* \|.
\end{equation}
Notice that for all $1 \le i,j \le m$, we have $U_i U_j^*=
\begin{pmatrix} C & -S \\ S & C \end{pmatrix},$ where
\begin{equation}
C= \cos(\frac{\pi(i-j)}{m}) I \quad {\rm and} \quad S=
\sin(\frac{\pi(i-j)}{m}) I .
\end{equation}
Thus, it will be enough to verify that the matrix
\begin{equation}
P + U_iU_j^*PU_jU_i^*= \begin{pmatrix} I + C^2 & -CS\\ -CS & S^2
\end{pmatrix}
\end{equation}
has norm strictly less than 2. However, this last matrix
decomposes into a direct sum of $2 \times 2$  Hermitian matrices
of the same form and one can check that the eigenvalues of such a
matrix are, $1 \pm \cos(\theta)$, with
$\theta=\frac{\pi(i-j)}{m}$. Thus, for $i \ne j, 1 \le i,j \le m$,
the angle $\theta$ is not an integer multiple of $\pi,$ so the
eigenvalues of these Hermitian matrices, and hence their norms,
will be bounded above by $2$. Indeed, a tight upper bound on the
norm is $1+ \cos(\frac{\pi}{m}).$
\end{ex}

\section{Uniformly weighted projections as optimal encoders for minimal decoherence}
\label{S:UWP}


The main theorem of this section characterizes optimal
encoding in our communication framework. We
prepare the result with three lemmas.

\begin{lemma}
Given a nonnegative operator $A$ with eigenvalues $\{\alpha_r\}
\subset [0,1]$ and at least one nonzero $\alpha_r$, then a vector
$\rho$ with $\rho_r \ge 0$, $\sum_r \rho_r=1$ maximizes
$$
   \sum_{r,s} ((1-\alpha_r) \alpha_s + (1- \alpha_s) \alpha_r )^2 \rho_r \rho_s
$$
if and only if at most two entries of $\rho$ are nonzero.
\end{lemma}
\begin{proof}
This is a standard Lagrange-multiplier argument. Suppose $\rho$ is
a maximizing vector and there is an index $r$ such that
$\rho_r=1$, then the assertion is true. On the other hand, assume
all $\rho_r < 1$. Then there exists a $\lambda \in \mathbb R$ such
that for any $r$ with $\rho_r \neq 0$,
$$
   \sum_s  ((1- \alpha_r) \alpha_s + (1- \alpha_s) \alpha_r )^2 \rho_s = \lambda \, .
$$
Since the left hand side is quadratic in $\alpha_r$, there can be
at most two different values for $\alpha_r$ that satisfy this
equation. Consequently, $\rho_r$ can be nonzero for at most two
$r$.
\end{proof}

\begin{lemma}
Let $A$ be a nonnegative operator with eigenvalues
$\{\alpha_r\}\subseteq [0,1]$, the maximum of which is given by
$\kappa=\|A\|$. Let the function $g: [0,\kappa]^2 \to \mathbb R$
be defined by $g(x,y) = ((1-kx)y + (1-ky)x)^2$. Then the maximum
of
$$
   G(p,x,y) = p^2 g(x,x) + 2 p (1-p) g(x,y) + (1-p)^2 g(y,y) \,
$$
over $p\in [0,1]$ and $\{(x,y):$ $\kappa \ge x\ge y\ge 0\}$ is
achieved for some $p$ when $x=\kappa$ and $y=0$.
\end{lemma}
\begin{proof}
We note that the auxiliary function
$$
   h(x,y) = x+y - 2 x y
$$
is harmonic in ${\mathbb D} = [0, \kappa]^2$ and piecewise linear
when restricted to the boundary $\partial \mathbb D$. Denote
$$
  H(p,x,y) = p^2 h (x,x) + 2 p (1-p) h(x,y) + (1-p)^2 h(y,y) \, .
$$
Let $\eta_{x,y}$ be the so-called harmonic measure on $\partial
\mathbb D$ for the point $(x,y)$. That is, for any harmonic
function $u$ that is continuous on $\mathbb D$, it gives
$u(x,y)=\int_{\partial \mathbb D} u \,d\eta_{x,y}$. Using a
suitable convex combination of harmonic measures for the points
$(x,x), (x,y), (y,x)$ and $(y,y)$, we construct $\mu_{p,x,y}$
supported on the boundary $\partial \mathbb D$ such that
$\int_{\partial \mathbb D} h\, d\mu_{p,x,y} = H(p,x,y)$. Since $h$
is piecewise linear on $\partial \mathbb D$, this measure can be
replaced by weights $w_{11}, w_{12}, w_{21}, w_{22} \ge 0$ at the
corners. These weights give
$$
   w_{11} h(\kappa,\kappa) + w_{12} h(\kappa,0) + w_{21} h(0,\kappa) + w_{22} h(0,0) = H(p,x,y) \,
$$
and can be determined as
$$
    w_{11} = \frac{1}{\kappa^2} (px +(1-p) y)^2 \, ,
$$
$$
    w_{22} = \frac{1}{\kappa^2} (p(\kappa-x) + (1-p)(\kappa-y))^2 \, ,
$$
and
\begin{eqnarray*}
    w_{12}=w_{21} &=& \frac{1}{\kappa^2}
     \big[ p^2x(\kappa-x) + p(1-p) (y
     (\kappa-x)+x(\kappa-y))\\
    & & + (1-p)^2 y(\kappa-y)\big] \, .
\end{eqnarray*}
It can be verified that $w_{12} = \sqrt{w_{11} w_{22}}$ and that
$w_{11}+2w_{12}+w_{22}=1$.

We observe that $g=h^2$ is sub-harmonic. Consequently, Jensen's
inequality yields
$$
  G(p,x,y) \leq \int_{\partial \mathbb D} g \, d\mu_{p,x,y}
$$
and since the restriction of $g$ to each side of $\partial \mathbb
D$ is convex, applying Jensen's inequality again gives us
\begin{align*}
   G(p,x,y) & \leq
   w_{11} g(\kappa,\kappa) + w_{12} g(\kappa,0) + w_{21} g(0,\kappa) + w_{22} g(0,0) \\
 &  = G(\sqrt{w_{11}},\kappa,0) \, .
\end{align*}
Thus, $G$ assumes its maximum when the point $(x,y)$ is on the
boundary of the square $\mathbb D$.
\end{proof}

\begin{lemma}\label{thm:DE}
Let $\Phi: B(\CC^d) \to B(\CC^m)\otimes B(\CC^l)$ be a
$C^*$-encoding and let $\{V_j\}_{j=1}^m$ be the associated coordinate
operators of rank $l<d$ 
as defined in Lemma~\ref{lem:coordop}. Let $Q_n= E_{n n}
\otimes I$, for $1\leq n \leq m$, be the orthogonal projection
onto the $n$th copy of ${\mathbb C}^l$ in $\CC^m\otimes \CC^l=
\oplus_{j=1}^m \CC^l$, and denote $\mathcal{E}_n$ the projective
measurement applied to any state $M \in B(\oplus_{j=1}^m \CC^l)$,
$$
                            \mathcal{E}_n(M) = Q_n M Q_n + Q_n^\perp M Q_n ^\perp \, .
$$
Then the maximum of the Hilbert-Schmidt norm $(\tr[Y Y^*])^{1/2}$
of the reconstruction error $Y=W-\Phi^*( \E_n (\Phi(W)))$ over all
states $W$ only depends on the operator norm $k_n =\|V_n^* V_n
\|$.
Specifically, we have
\begin{eqnarray}
   e(\Phi,\mathcal{E}_n)= \max_W (\tr[Y Y^*])^{1/2} = 2\ks_n (1-\ks_n)  \quad
{\rm when} \quad \ks_n \leq 1/2,
\end{eqnarray}
and
\begin{eqnarray}
   e(\Phi,\mathcal{E}_n)=\frac{\ks_n}{\sqrt{-2+8\ks_n-4\ks^2_n}} \quad {\rm
when} \quad \ks_n> 1/2.
\end{eqnarray}
\end{lemma}

\begin{proof}
We fix $n$ and abbreviate $A =A_n= V_n^* V_n$. After decoding, the
resulting error in the reconstructed state is
$$
    Y= (I-A)W A +  A W(I- A) \, .
$$
The square of the Hilbert-Schmidt norm is
\begin{align*}
   \tr [Y Y^*] &= \tr[ ((I- A)W A +  A W(I- A))^2 ]\\
     & = 2 \tr[ ((I- A)AW)^2 + (I- A)^2 W A^2 W] \, .
\end{align*}
In the eigenbasis of the operator $A$, we can express this as
$$
   \tr [ Y Y^*] = 2 \sum_{r,s}\left( (1-\alpha_r) \alpha_r (1- \alpha_s) \alpha_s |W_{r,s}|^2
                                         + (1-\alpha_r)^2 \alpha_s^2 |W_{r,s}|^2 \right) \, ,
$$
where the matrix with entries $(W_{r,s})$ represents the state $W$
in the eigenbasis of~$A$ and $\{\alpha_r\}$ are the corresponding
eigenvalues of~$A$.

We notice that if we assume $\alpha_r \leq 1$ for all $r$, the
expression for $\tr[Y Y^*]$ is just the square of a weighted
$\ell^2$-norm on the set of all states, and thus convex in $W$.
Moreover, the set of all states is convex, and therefore, a state
that maximizes the Hilbert-Schmidt norm of $Y$ must be a pure
state, $W_{r,s}=\phi_r \phi^*_s$, with $\sum_r |\phi_r|^2 =1$.
Since $|W_{r,s}|$ is symmetric with respect to exchanging $r$ and
$s$, we can symmetrize the term $(1-\alpha_r)^2 \alpha_s^2$ in the
above sum.

To determine a maximizer, we use a variational principle. We
replace the discrete spectrum of $A$ with a continuous variable
and examine the symmetrized quantity
\begin{align*}
   g(x,y) &= xy(1- x)(1- y) + \frac 1 2 (1- x)^2 y^2 + \frac 1 2 (1- y)^2 x^2 \, \\
          &= \frac 1 2 ((1- x)y + (1- y)x)^2 \, .
\end{align*}
A critical point of $g$ satisfies
$$
   (1-2 y)(y(1- x)+x(1- y))=0
$$
and the same equation with $x$ and $y$ exchanged. Thus, either
$x=y=0$ or $x=y=\frac{1}{2}$.


We denote $\kappa = \|A\|$.

{\noindent}\textit{Case I: $\kappa \leq 1/2$.} Considering the
boundaries, we have $g(x,0)\leq g(\kappa,0)=\frac 1 2 \kappa^2$
and $g(x,\kappa)=x(1-x)\kappa(1- \kappa) + \frac 1 2 (1- x)^2
\kappa^2
      + \frac 1 2 (1-\kappa)^2 x^2$ for $0 \leq x \leq \kappa$,
which is convex if $1-2\kappa \neq 0$ and linear otherwise. In
both cases the maximum on the boundary is either $g(0,\kappa)=
\frac 1 2 \kappa^2$ or $g(\kappa,\kappa)= 2 (1 - \kappa)^2
\kappa^2$. When $\kappa \leq \frac 1 2$, the latter expression is
the maximum. Moreover, the critical point is not contained in the
open square $(0, \kappa)^2$.

This means that the state which maximizes
$$\tr[Y Y^*] = 2  \sum_{r,s} ((1-\alpha_r) \alpha_r (1-\alpha_s) \alpha_s
+ (1-\alpha_r)^2 \alpha_s^2) |\phi_r|^2 |\phi_s|^2 $$ is given by
$\phi$ being an eigenstate of $A$ with highest eigenvalue
$\kappa$, and
$$
  \max_W \tr[ Y Y^*] = 4 (1 - \kappa)^2 \kappa^2 \, .
$$

{\noindent}\textit{Case II: $1/2 < \kappa \leq 1$.} Since the rank
of $A$ is not maximal, we know it has an eigenvalue zero. From the
preceding lemmas, we know that the probability vector
$\rho=|\phi|^2$ which maximizes
$$
   \tr[ Y Y^*] = \sum_{r,s} ((1-\alpha_r) \alpha_s + (1-\alpha_s) \alpha_r)^2 \rho_r \rho_s
$$
can have at most two nonzero entries, which correspond to the
maximal eigenvalue $\kappa$ and an eigenvalue zero.

Choosing $\rho$ accordingly, and optimizing
$$
  \tr[ Y Y^*] =  ( 4 p^2 (1-\kappa)^2\kappa^2 + 2p(1-p)\kappa^2 )
$$
gives
$$
  p= \frac{1}{2 - 4 (1-\kappa)^2} \, .
$$
Inserting this, we obtain
$$
    \max_W \tr[Y Y^*] = \frac{ \kappa^2 }{2-4(1- \kappa)^2} \, .
$$

After inspecting both cases for $\kappa$, we conclude that
\[
\max_W \tr[Y Y^*] = \left\{ \begin{array}{cl} 4(1- \kappa)^2
\kappa^2 \, ,&  \kappa\in [0, 1/2] \, ,\\ \frac{ \kappa^2
}{2-4(1-\kappa)^2} \, ,& \kappa \in (1/2,1]\, .
\end{array}\right. 
\]
\end{proof}

\begin{thm} \label{thm:cvx}
Consider the Hilbert space $\CC^m \otimes \CC^l$ and let
the set of generalized phase-damping channels 
$\mathcal{Q}\up 1$ be the convex hull of the channels
$\{\mathcal{E}_n\}_{n=1}^m$ defined with
the set of projections
$\{Q_n=E_{nn}\otimes I\}_{n=1}^m$. For any given $C^*$-encoding
$\Phi: B(\CC^d) \to B(\CC^m \otimes \CC^l)$, $l<d$, 
the worst case error is bounded below by
\begin{equation}\label{eq:errbd}
   \sup_{\mathcal{E} \in \mathcal{Q}\up 1} e(\Phi,\mathcal{E})
   \ge \left\{ \begin{array}{cl} 2(d-\frac{d^2}{ml}) \, ,& \frac{d}{ml}\in [0, 1/2]\, , \\
   \frac{ d
}{\sqrt{8 m l d - 2 m^2 l^2 - 4 d^2}} \, , & \frac{d}{ml} \in (1/2,1]\, ,
\end{array}\right. 
\end{equation}
and equality holds if and only if all
 $\{V_j^* V_j\}$ are uniformly weighted
projections; that is, for all $j\in \{1, 2, \dots m\}$, we have
$V_j^* V_j = k P_j$, with $P_j=P_j^* P_j$ and $k=\frac{d}{ml}$.
\end{thm}

\begin{proof}
Given a $C^*$-encoding $\Phi$, we optimize among
all convex combinations of $\mathcal{E}_n$
as defined in the preceding lemma.

We observe that for any probability vector $p \in \mathbb{R}_+^m$,
by Minkowski's inequality for the Hilbert-Schmidt norm
$$
   e(\Phi, \sum_{j} p_j \mathcal{E}_j)
   \leq  \sum_{j} p_j e(\Phi,\mathcal{E}_j) \, .
$$
The right-hand side is maximized by the probability vector
which is nonzero only in the index $j$ with the largest $e(\Phi,\mathcal{E}_j)$.
However, choosing the worst-case state for this channel
and the maximizing probability vector on the left-hand side
gives equality in the above inequality.

Therefore, optimal encoding suppresses the maximum
among all $e(\Phi,\mathcal{E}_n)$.

The function $e(\Phi,\mathcal{E}_n)$ is increasing in $k_n=\|A_n\|$. Since
we want to choose $\{A_j\}$ such that the maximum of $e(\Phi,\mathcal{E}_n)$
is minimized among all choices of $n$, we want to minimize
$\max_{1 \leq j \leq m} \|A_j\|$. The minimum is achieved when
equality holds in Inequality~(\ref{ineq:tr}). This requires that
the POVM $\{A_j\}$ consists of uniformly weighted projections.
\end{proof}

\begin{remark}
Recall that in Example~\ref{ex1} we saw how to construct uniformly
weighted rank-one projections resolving the identity. Then in
Example~\ref{ex2} we indicated how to generalize this construction
to higher-rank projections in the case that $l$ divides $d$. We
are aware of some other cases for which such families can be
constructed, but here we leave the general case as an open problem
that warrants further investigation.
\end{remark}

\section{Optimality for next to minimal decoherence}
\label{S:ONM}

Next, we investigate what happens if we choose $\Phi$ optimal
for minimal decoherence, and consider convex combinations of
measurements with
next to minimal impact. By the results of the preceding section,
we can assume that the coordinate operators $\{V_j\}_{j=1}^m$ of $\Phi$
factor a set of uniformly weighted projection operators,
$V_j^* V_j = k P_j$, with $k=d/ml$ and $\ks \sum_{j=1}^m P_j = I$.

\begin{defn}
Let $K$ be a subset of two elements from the index set $J=\{1, 2,
\dots m\}$. Define $Q_K=\sum_{j\in K} E_{jj}\otimes I$, and
let for $M \in B(\oplus_{j=1}^m \CC_l)$,
$$
  \mathcal{E}_K: M \mapsto Q_K M Q_K + Q_K^\perp M Q_K^\perp \, .
$$

Let $\mathcal{Q}\up 2$ be the convex hull of all transmission errors
$\mathcal{E}_K$ indexed by two-element
subsets $K$ of $J$.
\end{defn}

The following lemma reduces to the so-called Welch bound
in the special case that all projections $\{P_j\}$ are of rank $l=1$.

\begin{lemma}
Let $\{P_i\}_{i=1}^m$ be a set of rank-$l$ projections 
forming a uniformly 
weighted resolution of the identity $\frac{d}{ml}\sum_{j=1}^m P_j = I$
on 
a Hilbert space of dimension $d>l$. Then (a) if $d<2l$, we have that
for all pairs $i,j \in \{1, 2, \dots m\}$,  $\|P_i+P_j\|=2$; (b) 
if $d\ge 2l$, on the other hand, then
\begin{equation*}
\max_{i \ne j}\|P_i+P_j\| \ge 
                               1+ \sqrt{\frac{lm-d}{d(m-1)}}
\end{equation*}
and equality holds in this inequality if and only if for all $i,j \in \{1, 2, \dots m\}$,
$$
   \|P_i+P_j\| = 1+ \sqrt{\frac{lm-d}{d(m-1)}}\, .
$$
\end{lemma}
\begin{proof}
We can assume that the Hilbert space is $\CC^d$, equipped with the
canonical basis, and identify the projections with $d\times d$ matrices.

(a) We first show that in case $d< 2l$, the sum of two of the projections has 
norm $\|P_i+P_j\|=2$ for all $i, j \in J$.
To see this, we write the projections
as $2\times 2$ block matrices with the $(1,1)$-block being an $l\times l$ matrix, and 
choose a basis such that this block is the identity for
$P_i= \begin{pmatrix} I & 0\\0 & 0 \end{pmatrix}$.  Now when we 
compress $P_j$ to this first $l\times l$ block, we 
remove $r=d-l < l$ rows and 
columns and so by eigenvalue interlacing, the $l\times l$ block 
still has eigenvalue $1$ of multiplicity 
$l-r>0$. Consequently, $P_i + P_j$ has $2$ as an eigenvalue.

(b) We now consider the case $d\ge 2l$. 

Splitting $\CC^d= \CC^l \oplus \CC^l \oplus \CC^r$, $r\ge 0$,
we write $d\times d$ matrices as $3\times 3$ block 
matrices. By a suitable choice of basis, we 
may assume that $$P_1= \begin{pmatrix} I & 0 & 0\\0 & 0 & 0 
\\0 & 0 & 0 \end{pmatrix}\, .$$ In the same decomposition, each $P_j$ 
is written as a block matrix, and we call $B_j$ the $(1,1)$-block entry, so 
$B_j$ 
is an $l\times l$ matrix.

Now summing these matrices we have that $I + \sum_{j=2}^m B_j = 
\frac{lm}{d}I$ by the resolution of the identity and hence $\sum_{j=2}^m \tr(B_j) =\frac{l^2m}{d}-l$.
Thus, selecting for each $B_j$ its largest eigenvalue $\lambda_j$
gives the chain of inequalities
\begin{equation} \label{ineq:jref}
  \max_{j \ge 2} \lambda_j \ge \frac{1}{m-1}\sum_{j=2}^m \lambda_j \ge \frac{lm-d}{d(m-1)} \, .
\end{equation}


In the remaining part of the proof, we show that
$$
  \|P_1 + P_j \| = 1 + \sqrt{\lambda_j} \, .
$$
Once this is established, we have by the monotonicity of the square
root that
$$
  \max_{j \ge 2} \|P_1+P_j\| \ge 1 + \sqrt{\frac{lm-d}{d(m-1)}} \, 
$$
and equality implies that equality holds in the chain of Inequalities~(\ref{ineq:jref}),
which in turn gives that all $\lambda_j$ are equal, as well as
all $\|P_1 + P_j\|$.
Since the choice of $P_1$ is arbitrary, this argument shows that all pairs $\{P_i+P_j\}_{i \neq j}$
have equal norm.

For simplicity of notation, we continue the remaining part of the proof with $i=1$ and $j=2$.
We first show that up to unitary equivalence 
$P_2= \begin{pmatrix} D & R & 
0\\R & R_{22} &R_{23}\\0 & R_{32} & R_{33} \end{pmatrix}$,
with diagonal blocks $D$, $R$, and as yet unspecified parts $R_{22}, R_{23}, R_{32}$, and $R_{33}$.
To see this, first we write $P_2$ in $2\times 2$ block form with the 
$(1,1)$-block an $l\times l$ matrix and conjugate by 
a unitary to diagonalize $B_2$ as $D$.

Now we polar decompose the $(1,2)$-block as 
$RV^*$ where $R\ge 0$ is $l\times l$ and $V^*$ is $l\times(d-l)$, and by assumption $d-l \ge l$. 
Let $U$ be a 
$(d-l)\times(d-l)$ unitary so that $V^*U=(I\; 0)$. Conjugating by $ \begin{pmatrix} I & 0  
\\0 & U \end{pmatrix}$ we have that 
up to these two unitary transformations, $P_2= \begin{pmatrix} D & R & 0\\R & R_{22} & R_{23}\\0 
& R_{32} & R_{33} \end{pmatrix}$. But since $P_2$ is a projection, $D^2+R^2=D,$ and 
we see that $R$ is diagonal as well.  

If $R$ has entries that are zero, then the corresponding entries in $D$ must vanish also,
since by assumption $\|P_1+P_2\|<2$ which yields $\|D\|<1$. 
Again using that $P_2$ is a 
projection, $D R+ R R_{22}=R$ and hence $R_{22}$ is also diagonal
in the rows and columns where $R$ is non-zero. In addition, since 
$0= R R_{23}, R_{23}$ vanishes in those rows.  

We now observe, similarly as in Example~\ref{ex3}, that $P_2$ is a direct sum
of $2\times 2$ matrices and a piece which contains the blocks
that are not diagonal. 



If the eigenvector of $P_1+P_2$ corresponding to the largest eigenvalue 
had nonzero entries in
a row in which $D$ and $R$ vanish, then that eigenvalue would
be one. Indeed, this is the case if $B_2=D=R=0$, since then the ranges of $P_1$ and $P_2$ 
are orthogonal, and $\|P_1+P_2\|=1$.

However, if $D$ has rows with non-zero entries, then
we find the largest eigenvalue among the $2\times 2$ matrices.
%
To this end, 
pick a non-zero entry of $D$, say $\lambda$. Note that since $P_2$ is a projection, the 
corresponding diagonal entry of $R$ is then $\sqrt{\lambda-\lambda^2}$ and that 
of $R_{22}$ is $1-\lambda$.
Thus, in $P_1+P_2$ we have a $2\times 2$ matrix of the form $\begin{pmatrix} 1+\lambda & 
\sqrt{\lambda-\lambda^2}\\ \sqrt{\lambda-\lambda^2} & 1-\lambda \end{pmatrix}$.
One finds that the norm of this matrix is $1 + \sqrt{\lambda}$. Consequently,
the largest entry $\lambda_2$ in $D$ maximizes this norm, which shows that
$\| P_1+P_2\|=1 +\|\sqrt{B_2}\|$. 
\end{proof}

\begin{thm} \label{thm:SDE}
Assume the $C^*$-encoding $\Phi: B(\CC^d) \to B(\CC^m \otimes \CC^l)$ 
has coordinate operators $\{V_j\}_{j=1}^m$
with rank $l<\dd$, such that each $V_j^* V_j=kP_j$ with a projection
operator $P_j$ and $k=d/ml$. 
Denote
$$
  \kappa = \left\{\begin{array}{ll} 
                  \frac{2d}{ml} \, ,& d < 2l\\
                  \frac{d}{ml} + \sqrt{\frac{d(lm-d)}{m^2l^2(m-1)}}\, , & d \ge 2l\, , 
                 \end{array}\right.
$$
then the worst case error under the $C^*$-encoding $\Phi$ and transmission
channels from $\mathcal{Q}\up 2$ is bounded below by
$$
  \max_{\mathcal{E} \in \mathcal{Q}\up 2} e(\Phi,\mathcal{E})
  \ge \left\{\begin{array}{ll} 
               2\kappa (1-\kappa) \, ,& \kappa \in [0, 1/2]\\
                \frac{ \kappa
}{\sqrt{-2+8\kappa-4\kappa^2}} \, ,& \kappa \in (1/2,1]\, .
            \end{array}\right.  
$$
If $d \ge 2l$, then the value
$\max_{\mathcal{E} \in \mathcal{Q}\up 2} e(\Phi,\mathcal{E})$
attains the lower bound
if and only if for all $i,j \in \{1, 2, \dots m\}$,  
$$\|P_i+P_j\| = 1+ \sqrt{\frac{lm-d}{d(m-1)}}\, .$$
\end{thm}

\begin{proof}
Denote $A=\ks \sum_{j\in K} V_j^* V_j$. After decoding, the
resulting error in the reconstructed state is
$$
    Y= (I-A)WA + A W(I-A) \, .
$$
In the eigenbasis of the operator $A$, the square of the
Hilbert-Schmidt norm of $Y$ is
$$
   \tr [ Y Y^*] = 2 \sum_{r,s}\left( (1-\alpha_r) \alpha_r (1-\alpha_s) \alpha_s |W_{r,s}|^2
                                         + (1-\alpha_r)^2 \alpha_s^2 |W_{r,s}|^2 \right) \, ,
$$
where again the matrix with entries $(W_{r,s})$ represents the
state $W$ in the eigenbasis of~$A$ and $\{\alpha_r\}$ are the
corresponding eigenvalues of~$A$.

We notice that finding the state $W$ which maximizes the
worst-case error norm
is exactly the same problem as in Lemma~\ref{thm:DE} of the preceding section. 
By the monotonicity of the resulting expression in $\|A\|$
and the convexity of the Hilbert-Schmidt norm, we can
conclude analogously that the optimal encoding minimizes the
largest of the eigenvalues of all pairs $\{P_i+P_j\}_{i \neq j}$.
The preceding lemma states that this is the case if and only
if the norms of all these pairs are equal. The
common value of their norms then yields the claimed error bound.
\end{proof}

We describe a class of examples for which the lower bound
for the worst-case error is attained.

\begin{ex} \label{ex:tuf}
Let ${\mathbb C}^d = {\mathbb C}^q \otimes {\mathbb C}^l$, and
$\{P_j = \Pi_j \otimes I\}_{j=1}^m$ with rank-one projections
$\{\Pi_j\}$ giving a uniformly weighted resolution of the identity
on the first component  $ {\mathbb C}^q$. We follow ideas of
Holmes and Paulsen \cite{HP04} to characterize optimal uniformly
weighted projections of this type.

Let $\{f_j\}_{j=1}^m$ be a set of non-zero vectors, such that each
$f_j$ is contained in the span of the corresponding rank-one
projection $\Pi_j$, with the normalization $\|f_j\|= \sqrt{q/m}$.
One can check that this set is a Parseval frame.

The largest eigenvalue of $P_i + P_j$ is $\| P_i + P_j \| = \| 
 (\Pi_i + \Pi_j) \otimes I \| =\| \Pi_i + \Pi_j \|$. From identities
for the trace of the Grammian $G$ of the Parseval frame $\{f_j\}$,
we have $q=\tr G = \sum_{j=1}^m \|f_j\|^2$ and  $q=\tr G^* G =
\sum_{i,j} | \langle f_j , f_j \rangle |^2$, and we can infer
the usual form of the Welch bound \cite{Wel74,HP04,XZG05}
$$
  \max_{i\neq j} \|\Pi_i + \Pi_j \|
  = \max_{i \neq j} \big\{1 + \frac{m}{q} | \langle f_i, f_j \rangle | \big\}
\ge 1 + \sqrt{\frac{m-q}{q(m-1)}} \,
$$
and equality holds if and only if $|\langle f_i, f_j \rangle | =
\sqrt{q(m-q)/(m^2(m-1))}$ for all $i \neq j$.

Therefore, an optimal set of projections of the form $\{ \Pi_j
\otimes I\}$ with rank-one projections $\{\Pi_j\}$ is given by a
so-called two-uniform frame $\{f_j\}$ with all inner products
between two different frame vectors having the same absolute
value.
\end{ex}

\begin{remark}\label{rem:gen}
Now we briefly remark how to generalize to the case when $\mathop{\mathrm{dim}}{\mathcal
V} = s > 0 $. This includes the case $m=1$ of so-called decoherence-free subspaces.
The minimal error model is the convex hull of transmission
channels given by projections of both types
$\{Q_j = (E_{jj} \otimes I) \oplus 0\}_{j=1}^m$ or $\{Q_{m+n} = (0 \otimes 0) \oplus
E_{nn}\}_{n=1}^s$.

Suppose we have a $C^*$-encoding 
$\Phi: B(\CC^d) \to B((\CC^m \otimes \CC^l) \oplus \mathcal{V})$.
Then we can view identify the tensor product
as a direct sum, $(\CC^m \otimes \CC^l) \oplus \mathcal{V}
= (\oplus_{j=1}^m \CC^l) \oplus \mathcal V$ and
split the matrix $\tilde V$ associated with $\Phi$ similarly as in 
the proof of Lemma~\ref{lem:coordop} into
coordinate operators $\{V_j\}_{j=1}^m$ of rank $l$ and
$\{V_j\}_{j=m+1}^{m+s}$ of rank one.

We note that the coordinate operators give rise to a
POVM $\{A_j\}_{j=1}^{m+s}$
with $m$ rank-$l$ operators $\{A_j\}_{j=1}^m$ and $s$ rank-one
operators $\{A_j\}_{j=m+1}^{m+s}$.

In analogy with Theorem~\ref{thm:cvx}, we can then derive a modified
version of the lower bound for the worst case error, which
amounts to replacing $ml$ with $ml+s$ in Ineq.~(\ref{eq:errbd}).
Again, equality would hold for $\Phi$ with coordinate
operators that factor uniformly weighted projections.
However, apart from the case when $s$ divides $l$ it is not clear 
under which general conditions an optimal $\Phi$ can be constructed. 

Similarly, to find the best $\Phi$ for the next to minimal error model, 
we require the maximal eigenvalue among all $\{P_i + P_j\}_{i\neq j}$ 
to be minimized, but at this point it is not clear whether there are enough
interesting examples of this type.
\end{remark}

\noindent{\it Acknowledgments.} We would like to thank the hospitality
of the Banff International Research Station, where some of this work took
shape. We would also like to thank Mary Beth Ruskai for valuable criticism
that improved this paper and C\'edric B\'eny for helpful discussions. This work
was partially supported by NSERC and NSF.



\begin{thebibliography}{30}
\expandafter\ifx\csname
natexlab\endcsname\relax\def\natexlab#1{#1}\fi
\expandafter\ifx\csname bibnamefont\endcsname\relax
  \def\bibnamefont#1{#1}\fi
\expandafter\ifx\csname bibfnamefont\endcsname\relax
  \def\bibfnamefont#1{#1}\fi
\expandafter\ifx\csname citenamefont\endcsname\relax
  \def\citenamefont#1{#1}\fi
\expandafter\ifx\csname url\endcsname\relax
  \def\url#1{\texttt{#1}}\fi
\expandafter\ifx\csname
urlprefix\endcsname\relax\def\urlprefix{URL }\fi
\providecommand{\bibinfo}[2]{#2}
\providecommand{\eprint}[2][]{\url{#2}}



\bibitem{PSE96}
\bibinfo{author}{\bibfnamefont{G. M.}~\bibnamefont{Palma}},
\bibinfo{author}{\bibfnamefont{K.-A.}~\bibnamefont{Suominen}} \bibnamefont{and}
  \bibinfo{author}{\bibfnamefont{A.}~\bibnamefont{Ekert}},
\textit{Quantum computers and dissipation},
\bibinfo{journal}{Proc. Royal Soc. A}
\textbf{\bibinfo{volume}{452}},
  \bibinfo{pages}{567-584} (\bibinfo{year}{1996}).

\bibitem{DG97c}
\bibinfo{author}{\bibfnamefont{L.-M.} \bibnamefont{Duan}} \bibnamefont{and}
  \bibinfo{author}{\bibfnamefont{G.-C.} \bibnamefont{Guo}},
 \textit{Preserving coherence in quantum computation by pairing
quantum bits}, \bibinfo{journal}{Phys. Rev. Lett.}
\textbf{\bibinfo{volume}{79}},
  \bibinfo{pages}{1953-1956} (\bibinfo{year}{1997}).

\bibitem{ZR97c}
\bibinfo{author}{\bibfnamefont{P.}~\bibnamefont{Zanardi}} \bibnamefont{and}
  \bibinfo{author}{\bibfnamefont{M.}~\bibnamefont{Rasetti}},
 \textit{Noiseless quantum codes}, \bibinfo{journal}{Phys. Rev. Lett.} \textbf{\bibinfo{volume}{79}},
  \bibinfo{pages}{3306-3309} (\bibinfo{year}{1997}).

\bibitem{LCW98a}
\bibinfo{author}{\bibfnamefont{D. A.}~\bibnamefont{Lidar}},
  \bibinfo{author}{\bibfnamefont{I. L.}~\bibnamefont{Chuang}}, \bibnamefont{and}
  \bibinfo{author}{\bibfnamefont{K. B.}~\bibnamefont{Whaley}},
\textit{Decoherence free subspaces for quantum computation},
\bibinfo{journal}{Phys. Rev. Lett.} \textbf{\bibinfo{volume}{81}},
  \bibinfo{pages}{2594-2597} (\bibinfo{year}{1998}).

\bibitem{KLV00a}
\bibinfo{author}{\bibfnamefont{E.}~\bibnamefont{Knill}},
  \bibinfo{author}{\bibfnamefont{R.}~\bibnamefont{Laflamme}}, \bibnamefont{and}
  \bibinfo{author}{\bibfnamefont{L.}~\bibnamefont{Viola}},
\textit{Theory of quantum error correction for general noise},
\bibinfo{journal}{Phys. Rev. Lett.} \textbf{\bibinfo{volume}{84}},
  \bibinfo{pages}{2525-2528} (\bibinfo{year}{2000}).

\bibitem{Zan01b}
\bibinfo{author}{\bibfnamefont{P.}~\bibnamefont{Zanardi}},
\textit{Stabilizing quantum information},  \bibinfo{journal}{Phys.
Rev. A} \textbf{\bibinfo{volume}{63}},
  \bibinfo{pages}{12301/1-4} (\bibinfo{year}{2001}).

\bibitem{KBLW01a}
\bibinfo{author}{\bibfnamefont{J.}~\bibnamefont{Kempe}},
  \bibinfo{author}{\bibfnamefont{D.}~\bibnamefont{Bacon}},
  \bibinfo{author}{\bibfnamefont{D.~A.} \bibnamefont{Lidar}}, \bibnamefont{and}
  \bibinfo{author}{\bibfnamefont{K.~B.} \bibnamefont{Whaley}},
 \textit{Theory of decoherence-free fault-tolerant universal
quantum computation,} \bibinfo{journal}{Phys. Rev. A}
\textbf{\bibinfo{volume}{63}},
  \bibinfo{pages}{42307/1-29} (\bibinfo{year}{2001}).

\bibitem{KBAW00}
\bibinfo{author}{\bibfnamefont{P.~G.}~\bibnamefont{Kwiat}},
  \bibinfo{author}{\bibfnamefont{A.~J.}~\bibnamefont{Berglund}},
  \bibinfo{author}{\bibfnamefont{J.~B.} \bibnamefont{Altepeter}}, \bibnamefont{and}
  \bibinfo{author}{\bibfnamefont{A.~G.} \bibnamefont{White}},
  \textit{A decoherence-free quantum memory using trapped ions},
  \bibinfo{journal}{Science} \textbf{\bibinfo{volume}{290}},
  \bibinfo{pages}{498-501} (\bibinfo{year}{2000}).

\bibitem{KMRSIMW01}
\bibinfo{author}{\bibfnamefont{D.}~\bibnamefont{Kielpinski}},
  \bibinfo{author}{\bibfnamefont{V.}~\bibnamefont{Meyer}},
  \bibinfo{author}{\bibfnamefont{M.~A.} \bibnamefont{Rowe}},
\bibinfo{author}{\bibfnamefont{C.~A.} \bibnamefont{Sackett}},
\bibinfo{author}{\bibfnamefont{W.~M.} \bibnamefont{Itano}},
\bibinfo{author}{\bibfnamefont{C.} \bibnamefont{Monroe}},
  \bibnamefont{and}
  \bibinfo{author}{\bibfnamefont{D.~J.} \bibnamefont{Wineland}},
\textit{Experimental verification of decoherence-free subspaces},
\bibinfo{journal}{Science} \textbf{\bibinfo{volume}{291}},
  \bibinfo{pages}{1013-1015} (\bibinfo{year}{2001}).

\bibitem{FVHTC02}
\bibinfo{author}{\bibfnamefont{E.~M.}~\bibnamefont{Fortunato}},
  \bibinfo{author}{\bibfnamefont{L.}~\bibnamefont{Viola}},
  \bibinfo{author}{\bibfnamefont{J.} \bibnamefont{Hodges}},
\bibinfo{author}{\bibfnamefont{G.} \bibnamefont{Teklemariam}}, \bibnamefont{and}
\bibinfo{author}{\bibfnamefont{D.~G.} \bibnamefont{Cory}},
\textit{Implementation of universal control on a decoherence-free
qubit},
\bibinfo{journal}{New J. Phys.} \textbf{\bibinfo{volume}{4}},
  \bibinfo{pages}{5.1-5.20} (\bibinfo{year}{2002}).

\bibitem{VFPKLC03}
  \bibinfo{author}{\bibfnamefont{L.}~\bibnamefont{Viola}},
  \bibinfo{author}{\bibfnamefont{E.~M.}~\bibnamefont{Fortunato}},
  \bibinfo{author}{\bibfnamefont{M.~A.} \bibnamefont{Pravia}},
\bibinfo{author}{\bibfnamefont{E.} \bibnamefont{Knill}},
\bibinfo{author}{\bibfnamefont{R.} \bibnamefont{Laflamme}},  \bibnamefont{and}
\bibinfo{author}{\bibfnamefont{D.~G.} \bibnamefont{Cory}},
\textit{Experimental realization of noiseless subsystems for
quantum information processing,}  \bibinfo{journal}{Science}
\textbf{\bibinfo{volume}{293}},
  \bibinfo{pages}{2059-2063} (\bibinfo{year}{2003}).


\bibitem{BRS03}
\bibinfo{author}{\bibfnamefont{S.~D.}~\bibnamefont{Bartlett}},
  \bibinfo{author}{\bibfnamefont{T.}~\bibnamefont{Rudolph}} \bibnamefont{and}
  \bibinfo{author}{\bibfnamefont{R.~W.}~\bibnamefont{Spekkens}},
\textit{Classical and quantum communication without a shared
reference frame},  \bibinfo{journal}{Phys. Rev.
Lett.},\textbf{\bibinfo{volume}{91}},
  \bibinfo{pages}{027901/1-4} (\bibinfo{year}{2003}).

\bibitem{BGLPS04}
  \bibinfo{author}{\bibfnamefont{J.~-C.}~\bibnamefont{Boileau}},
  \bibinfo{author}{\bibfnamefont{D.}~\bibnamefont{Gottesman}},
  \bibinfo{author}{\bibfnamefont{M.~A.} \bibnamefont{Pravia}},
\bibinfo{author}{\bibfnamefont{R.} \bibnamefont{Laflamme}},
\bibinfo{author}{\bibfnamefont{D.} \bibnamefont{Poulin}},  \bibnamefont{and}
\bibinfo{author}{\bibfnamefont{R.~W.} \bibnamefont{Spekkens}},
\textit{Robust polarization-based quantum key distribution over
collective-noise channel},  \bibinfo{journal}{Phys. Rev. Lett.}
\textbf{\bibinfo{volume}{92}},
  \bibinfo{pages}{17901/1-4} (\bibinfo{year}{2004}).


\bibitem{WZL05}
\bibinfo{author}{\bibfnamefont{L.~A.}~\bibnamefont{Wu}},
  \bibinfo{author}{\bibfnamefont{P.}~\bibnamefont{Zanardi}} \bibnamefont{and}
  \bibinfo{author}{\bibfnamefont{D.~A.}~\bibnamefont{Lidar}},
\textit{Holonomic quantum computation in decoherence-free
subspaces},
\bibinfo{journal}{Phys. Rev. Lett.},\textbf{\bibinfo{volume}{95}},
  \bibinfo{pages}{130501/1-4} (\bibinfo{year}{2005}).

\bibitem{DMS04}
\bibinfo{author}{\bibfnamefont{O.}~\bibnamefont{Dreyer}},
\bibinfo{author}{\bibfnamefont{F.}~\bibnamefont{Markopoulou}},
 \bibnamefont{and}
  \bibinfo{author}{\bibfnamefont{L.}~\bibnamefont{Smolin}},
 \textit{Symmetry and entropy of black hole horizons},  \bibinfo{journal}{E-print: arxiv.org/hep-th/0409056}.

\bibitem{KM05}
\bibinfo{author}{\bibfnamefont{D.~W.}~\bibnamefont{Kribs}},
 \bibnamefont{and}
\bibinfo{author}{\bibfnamefont{F.}~\bibnamefont{Markopoulou}},
  \emph{\bibinfo{title}{Geometry from quantum particles,}} \bibinfo{journal}{E-print: arxiv.org/gr-qc/0510052}.

\bibitem{BMS06}
\bibinfo{author}{\bibfnamefont{S.~O.}~\bibnamefont{Bilson-Thompson}},
\bibinfo{author}{\bibfnamefont{F.}~\bibnamefont{Markopoulou}},
 \bibnamefont{and}
  \bibinfo{author}{\bibfnamefont{L.}~\bibnamefont{Smolin}},
 \textit{Quantum gravity and the standard model},  \bibinfo{journal}{E-print: arxiv.org/hep-th/0603022}.


\bibitem{KLP05}
\bibinfo{author}{\bibfnamefont{D.~W.}~\bibnamefont{Kribs}},
  \bibinfo{author}{\bibfnamefont{R.}~\bibnamefont{Laflamme}} \bibnamefont{and}
  \bibinfo{author}{\bibfnamefont{D.}~\bibnamefont{Poulin}},
  \emph{\bibinfo{title}{Unified and generalized approach to quantum error correction,}} \bibinfo{journal}{Phys. Rev. Lett.},\textbf{\bibinfo{volume}{94}},
  \bibinfo{pages}{180501/1-4} (\bibinfo{year}{2005}).


\bibitem{KLPL06}
\bibinfo{author}{\bibfnamefont{D.~W.}~\bibnamefont{Kribs}},
  \bibinfo{author}{\bibfnamefont{R.}~\bibnamefont{Laflamme}},   \bibinfo{author}{\bibfnamefont{D.}~\bibnamefont{Poulin}}, \bibnamefont{and}
  \bibinfo{author}{\bibfnamefont{M.}~\bibnamefont{Lesosky}},
  \emph{\bibinfo{title}{Operator quantum error correction,}} \bibinfo{journal}{Quantum Inf. \& Comp., to appear.}

\bibitem{LS05}
\bibinfo{author}{\bibfnamefont{D.~A.}~\bibnamefont{Lidar}} \bibnamefont{and}
\bibinfo{author}{\bibfnamefont{A.}~\bibnamefont{Shabani}},
 \textit{Theory of initialization-free decoherence-free subspaces and subsystems},
\bibinfo{journal}{Phys. Rev. A},\textbf{\bibinfo{volume}{72}},
  \bibinfo{pages}{042303/1-14} (\bibinfo{year}{2005}).

\bibitem{NP05}
\bibinfo{author}{\bibfnamefont{M.~A.}~\bibnamefont{Nielsen}} \bibnamefont{and}
  \bibinfo{author}{\bibfnamefont{D.}~\bibnamefont{Poulin}},
 \textit{Algebraic and information-theoretic conditions for operator quantum error correction},
 \bibinfo{journal}{E-print: arXiv.org: quant-ph/0506069}.

\bibitem{CK06}
\bibinfo{author}{\bibfnamefont{D.~W.}~\bibnamefont{Kribs}}, \bibnamefont{and}
  \bibinfo{author}{\bibfnamefont{M.~D.}~\bibnamefont{Choi}},
   \emph{\bibinfo{title}{Method to find quantum noiseless subsystems,}} \bibinfo{journal}{Phys. Rev. Lett.},\textbf{\bibinfo{volume}{96}},
  \bibinfo{pages}{050501/1-4} (\bibinfo{year}{2006}).

\bibitem{Kni06a}
\bibinfo{author}{\bibfnamefont{E.}~\bibnamefont{Knill}},
\textit{On protected realizations of quantum information},
\bibinfo{journal}{E-print: arXiv.org: quant-ph/0603252}.






\bibitem{SH03}
\bibinfo{author}{Th. Strohmer} \bibnamefont{and}
\bibinfo{author}{R.~W. Heath, Jr.},
\bibinfo{title}{{\it Grassmannian frames with applications to coding and communication}},
 \bibinfo{journal}{Appl. Comput. Harmon. Anal.},
  \bibinfo{volume}{{\bf 14}},
    \bibinfo{year}{2003},
     \bibinfo{number}{(3)},
  \bibinfo{pages}{257--275}.

  \bibitem{HP04}
\bibinfo{author}{R.~B. Holmes}  \bibnamefont{and}
\bibinfo{author}{V.~I. Paulsen},
\bibinfo{title}{{\it Optimal frames for erasures}},
\bibinfo{journal}{Linear Algebra Appl.},
 \bibinfo{volume}{{\bf 377}},
  \bibinfo{year}{2004},
  \bibinfo{pages}{31--51}.



  \bibitem{BP05}
\bibinfo{author}{B.~G. Bodmann, }
\bibnamefont{and}
\bibinfo{author}{V.~I. Paulsen},
\bibinfo{title}{{\it Frames, graphs and erasures}},
\bibinfo{journal}{Linear Algebra Appl.},
 \bibinfo{volume}{{\bf 404}},
  \bibinfo{year}{2005},
  \bibinfo{pages}{118--146}.

\bibitem{XZG05}
\bibinfo{author}{P. Xia},
\bibinfo{author}{Sh. Zhou} \bibnamefont{and}
\bibinfo{author}{G.~B. Giannakis},
\bibinfo{title}{{\it Achieving the Welch Bound with Difference Sets}},
 \bibinfo{journal}{IEEE Transactions on Information Theory},
 \bibinfo{volume}{{\bf 51}}
 \bibinfo{number}{(5)},
 \bibinfo{year}{2005},
\bibinfo{pages}{1900-1907}.

\bibitem{Kal06}
\bibinfo{author}{D. Kalra},
\bibinfo{title}{{\it Complex equiangular cyclic frames and erasures}},
\bibinfo{journal}{E-print: arXiv.org: math/0602342}.

\bibitem{NieChu00} M.~A. Nielsen, I.~L. Chuang,
\textit{Quantum computation and quantum information}, Cambridge
University Press, 2000.

\bibitem{Pau02} V. Paulsen,
\textit{Completely bounded maps and operator algebras,} Cambridge
University Press, Cambridge, United Kingdom, 2002.

\bibitem{BenZyc06} I. Bengtsson, K. Zyczkowski,
\textit{Geometry of quantum states}, Cambridge University Press,
2006.

\bibitem{Kup03}
G.\ Kuperberg, 
\textit{The capacity of hybrid quantum memory}, 
IEEE Trans.\ Inform.\ Theory {\bf 49} (6), 1465--1473 (2003).

\bibitem{Dav96} K.~R. Davidson,
\textit{$C^*$-algebras by example}, Fields Institute Monographs,
{\bf 6}, Amer.\ Math.\ Soc., Providence, 1996.



\bibitem{CK04}
\bibinfo{author}{P.~G. Casazza} \bibnamefont{and} \bibinfo{author}{G. Kutyniok},
\bibinfo{title}{{\it Frames of subspaces}},
\bibinfo{booktitle}{Wavelets, frames and operator theory},
\bibinfo{series}{Contemp. Math.},
 \bibinfo{volume}{{\bf 345}},
 \bibinfo{publisher}{Amer.\ Math.\ Soc.},
 \bibinfo{address}{Providence, RI},
  \bibinfo{year}{2004},
   \bibinfo{pages}{87--113}.

\bibitem{CKL06}
\bibinfo{author}{P.~G. Casazza},
\bibinfo{author}{G. Kutyniok} \bibnamefont{and}
\bibinfo{author}{S. Li},
\bibinfo{title}{{\it Fusion Frames and Distributed Processing}},
\bibinfo{year}{2006},
\bibinfo{publisher}{E-print: arXiv.org: math.FA/0605374}.



\bibitem{Wel74} L. Welch, \textit{Lower bounds on the maxiumum
cross correlation of signals}, IEEE Trans.\ Inform.\ Theory, {\bf IT-20} (3),
397--399 (1974).



\end{thebibliography}
\end{document}